\documentclass[preprint2]{aastex631}

\shorttitle{An explosive outflow in IRAS12326$-$6245}
\shortauthors{Zapata et al.}
\graphicspath{{./}{figures/}}
\usepackage{hyperref}
\newcommand{\dechms}[4]{$#1^{\rm h}#2^{\rm m}#3\mbox{$^{\rm s}\mskip-7.6mu.\,$}#4$}

\newcommand{\decdms}[4]{$-#1^{\circ}#2'#3\mbox{$''\mskip-7.6mu.\,$}#4$}
\usepackage{graphicx}

\begin{document}


\title{One, Two, Three ... An Explosive Outflow in IRAS 12326$-$6245 revealed by ALMA}

\author[0000-0003-2343-7937]{Luis A. Zapata}
\affiliation{Instituto de Radioastronom{\'i}a y Astrof{\'i}sica, Universidad Nacional Aut{\'o}noma de M{\'e}xico, 58090, Morelia, Michoac{\'a}n, M\'exico}

\author[0000-0001-5811-0454]{Manuel Fern\'andez-L\'opez}
\affiliation{Instituto Argentino de Radioastronomía (CCT-La Plata, CONICET; CICPBA), C.C. No. 5, 1894, Villa Elisa, Buenos Aires, Argentina}

\author[0000-0003-1014-3390]{Silvia Leurini}
\affiliation{INAF – Osservatorio Astronomico di Cagliari, Via della Scienza 5, 09047 Selargius (CA), Italy}

\author[0000-0003-2630-3774]{Estrella Guzmán Ccolque}
\affiliation{Instituto Argentino de Radioastronomía (CCT-La Plata, CONICET; CICPBA), C.C. No. 5, 1894, Villa Elisa, Buenos Aires, Argentina}

\author{Skretas, I. M.}
\affiliation{Max-Planck-Institut für Radioastronomie, Auf dem Hügel 69, 53121 Bonn, Germany}

\author[0000-0003-2343-7937]{Luis F. Rodr\' \i guez}
\affiliation{Instituto de Radioastronom{\'i}a y Astrof{\'i}sica, Universidad Nacional Aut{\'o}noma de M{\'e}xico, 58090, Morelia, Michoac{\'a}n, M\'exico}

\author[0000-0003-2343-7937]{Aina, Palau}
\affiliation{Instituto de Radioastronom{\'i}a y Astrof{\'i}sica, Universidad Nacional Aut{\'o}noma de M{\'e}xico, 58090, Morelia, Michoac{\'a}n, M\'exico}

\author[0000-0001-6459-0669]{Karl M. Menten}
\affiliation{Max-Planck-Institut für Radioastronomie, Auf dem Hügel 69, 53121 Bonn, Germany}

\author[0000-0003-4516-3981]{Friedrich, Wyrowski}
\affiliation{Max-Planck-Institut für Radioastronomie, Auf dem Hügel 69, 53121 Bonn, Germany}



\begin{abstract}
In the last years there has been a substantial increase in the number of the reported massive and luminous star-forming regions with related explosive outflows thanks 
to the superb sensitivity and angular resolution provided by the new radio, infrared, and optical facilities. Here, we report one more explosive outflow related with the massive and bright star-forming 
region IRAS 12326$-$6245 using Band 6 sensitive and high angular resolution ($\sim$0.2$''$) Atacama Large Millimeter/Submillimeter Array (ALMA) observations. We find over 10 
molecular and collimated well-defined streamers, with Hubble-Lemaitre like expansion motions, and pointing right to the center of a dusty and molecular shell (reported for the first time here)
localized in the northern part of the UCHII region known as G301.1A. The estimated kinematic age, and energy for the explosion are $\sim$700 yrs, and 10$^{48}$ erg, respectively.  
Taking into account the recently reported explosive outflows together with IRAS 12326$-$6245, we estimate an event rate of once every 90 yr in our Galaxy, similar to the formation rate of massive stars. 
\end{abstract}

\keywords{Interstellar molecules (849) --- Millimeter Astronomy (1736) --- Circumstellar gas (238) --- High resolution spectroscopy (2096)}


\section{Introduction} \label{sec:intro}

Explosive outflows are a relatively new class of molecular outflows related to star-forming regions. These outflows are probably 
associated with the disruption of a non-hierarchical massive and young stellar system (perhaps triggered by the merger of young 
massive stars), or with a protostellar collision \citep[e.g.,][]{2011Bally,2013Zapata,2016Bally,2017Bally,2009Zapata,2021Raga}. 
The explosive flows are impulsive, and possibly created by a single energetic and brief event \citep{2005Bally}. 
These flows consist of dozens of collimated CO streamers, [FeII] ``fingertips", and H$_2$ wakes pointing
back approximately to a central position \citep[see for instance the cases of Orion BN/KL, G5.89$-$0.39 and IRAS16076$-$5134][]
{1993Allen,2009Zapata,2016Bally,2020Zapata,2022GuzmanCcolque}, reminiscent of an explosive dispersal event \citep[e.g.,][]{2018Loiseau}. 
The CO streamers, tracing these outflows, are radially distributed, and appear nearly isotropic on the sky, presenting well defined Hubble-Lemaitre like 
velocity gradient along their length. This isotropic distribution makes the red- and blue-shifted streamers to appear intertwined, with the CO emission reaching
 velocity in the line wings of up to $100$ km s$^{-1}$. The energy of these events amounts for over $\sim10^{48}$~ergs, placing them among the most energetic 
 outflows in the Galaxy. In the case of Orion BN/KL (the nearest of all these sources) the dynamical ages of most CO streamers are close to 500 years, 
 exactly the same age of the disruption of a non-hierarchical massive and young stellar system \citep{2005Gomez, 2009Zapata}, which is the event that probably 
 originated the outflow. The disruption of the stellar cluster is clearly traced by the proper motions of at least four stellar objects that recede from 
 the same central position as the molecular gas  \citep{2020Rodriguez}. Orion BN/KL is the only case with data from the originating protostellar system, searches for which have so far been elusive
 in the rest of the cases for now \citep[DR21, G5.89$-$0.39, S106$-$IR, IRAS16076$-$5134:][]{2013Zapata,2020Zapata,2017Bally,2022GuzmanCcolque}, 
 where the evidence comes form the kinematics of molecular gas.
 
With the discovery of new explosive outflows associated with massive star-forming regions the estimated rate of events 
(1 every 100 years, approximately see Sect. 4) is comparable to the rate of supernovae, which indicates that dynamic interactions in highly 
clustered systems of protostars and even stellar collisions may be a common scenario during the first stages of massive star formation. 
However, because of the scarcity of these events and the observational challenges that they pose, little is known about the true nature
 of the engine of these outflows.

The luminous source IRAS12326$-$6245 (AGAL301.136$-$00.226 or G301.1364-00.2249) lies at the southern rim of the infrared bubble nebula S169: a semi-spherical 
structure of $\sim1.2$\,pc in radius, harbouring at least 10 early-type main sequence stars \citep{2021Duronea}. 
The usually adopted distance to S169 and IRAS12326$-$6245 of 4.4\,kpc \citep{1997Osterloh} has recently been revised \citep{2021Duronea}. 
They showed the association between S169 and the IRAS12326$-$6245 molecular condensation (named MC3 in their work), proposed a model for the kinematics of the S169 bubble, 
and estimated a new distance to both structures of 2.03$^{+0.77}_{-0.61}$\,kpc. In this work, we adopt the 2.03 kpc distance derived by \cite{2021Duronea}.
 We, accordingly, recalculate sizes, luminosities and other magnitudes extracted from the literature and affected by the change in the distance throughout the text.

The S169 bubble has a dynamical age of $10^5$\,yr 
and it is expanding at $\sim12$  km s$^{-1}$ \citep{2021Duronea}. Its southern rim is marked by more than 30 protostellar objects detected by MSX or 
Spitzer, product of a probable triggered star-formation process. It is delineated by a ridge of dense gas, and harbours a Herschel submillimeter dense 
dusty core \citep{2021Duronea}. The molecular ridge shows several condensations of gas, of which MC3 contains 5500 M$_\odot$ within an area of 0.94\,pc, 
and is very bright in different molecular tracers of dense gas \citep[CS, HNCO][]{1996Bronfman,2000Zinchenko}, dust continuum 
emission \citep[from 1.2\,mm to 0.010\,mm, e.g.,][]{2000Henning,2004Faundez}, and also in the centimeter regime \citep[e.g.,][]{1998Walsh,2007Urquhart}. 
Through higher angular resolution observations,  the MC3 molecular condensation is resolved into two MIR continuum sources \citep{2000Henning}, 
coincident with two UC\,HII regions \citep{1998Walsh}. The southern of these two sources matches the peak position of the IRAS12326$-$6245 source, with 
a luminosity of 8.1$\times10^4$ L$_\odot$ \citep{1995Zinchenko,1997Osterloh}.

In addition, IRAS12326$-$6245 is associated with various high-mass star formation indicators. It shows maser emission 
from CH$_3$OH, H$_2$O, and OH \citep[e.g.,][]{1998MacLeod}; single-dish observations show very broad radio recombination 
lines associated with the UC HII region \citep[$FWHM\sim68$ km s$^{-1}$ ][]{2005Araya}, and signs of active outflow activity \citep{2000Henning,2005Araya}.

In this work, we report interferometric CO observations taken with the Atacama Large Millimeter/Submillimeter Array (ALMA) observatory, which show, for the first time, 
the true nature of the main outflow at the heart of IRAS12326$-$6245. The ALMA data is part of a survey toward six sources with extremely high-velocity wings revealed 
in the APEX spectra, highly indicative of jets, and luminosities $>$10$^5$ L$_\odot$, corresponding to ZAMS stars of 20$-$40 M$_\odot$. These ALMA observations are 
expected to reveal the structure of the flows, and therefore are ideal to test model predictions, and verify whether jets exist in massive young stellar objects. 
As we will show, these ALMA data have also the capability to reveal the new kind of extreme molecular flows, known as the explosive outflows \citep{2009Zapata}.

We report the telescope setup and the process of image production in Section 2. 
In Section 3 we present the main results regarding the outflow activity in the region, and finally, Section 4 is devoted to show our main conclusions. 

\begin{figure*}
\centering
\includegraphics[scale=0.45]{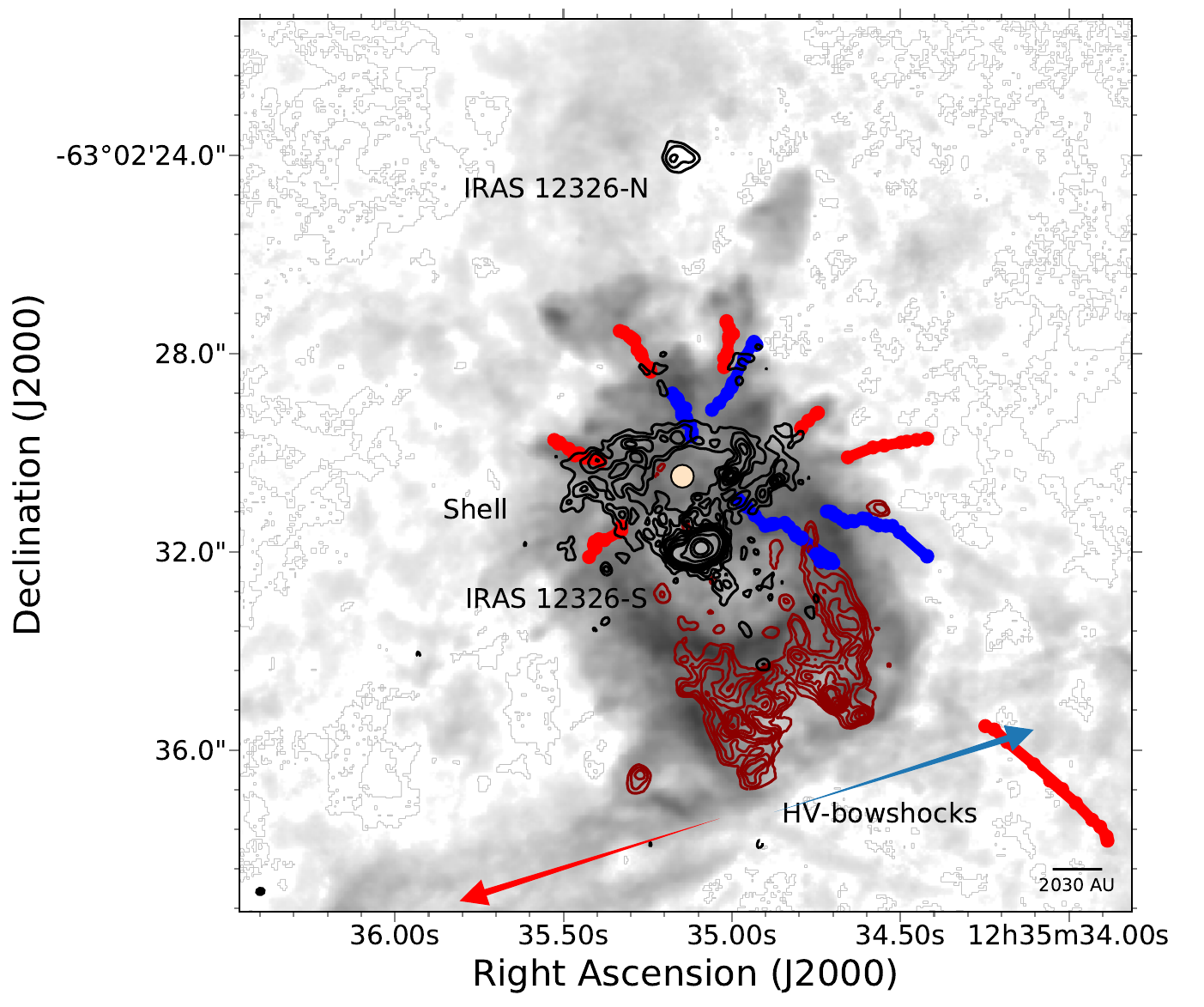}\\ \hspace{1.2cm}
\includegraphics[scale=0.45]{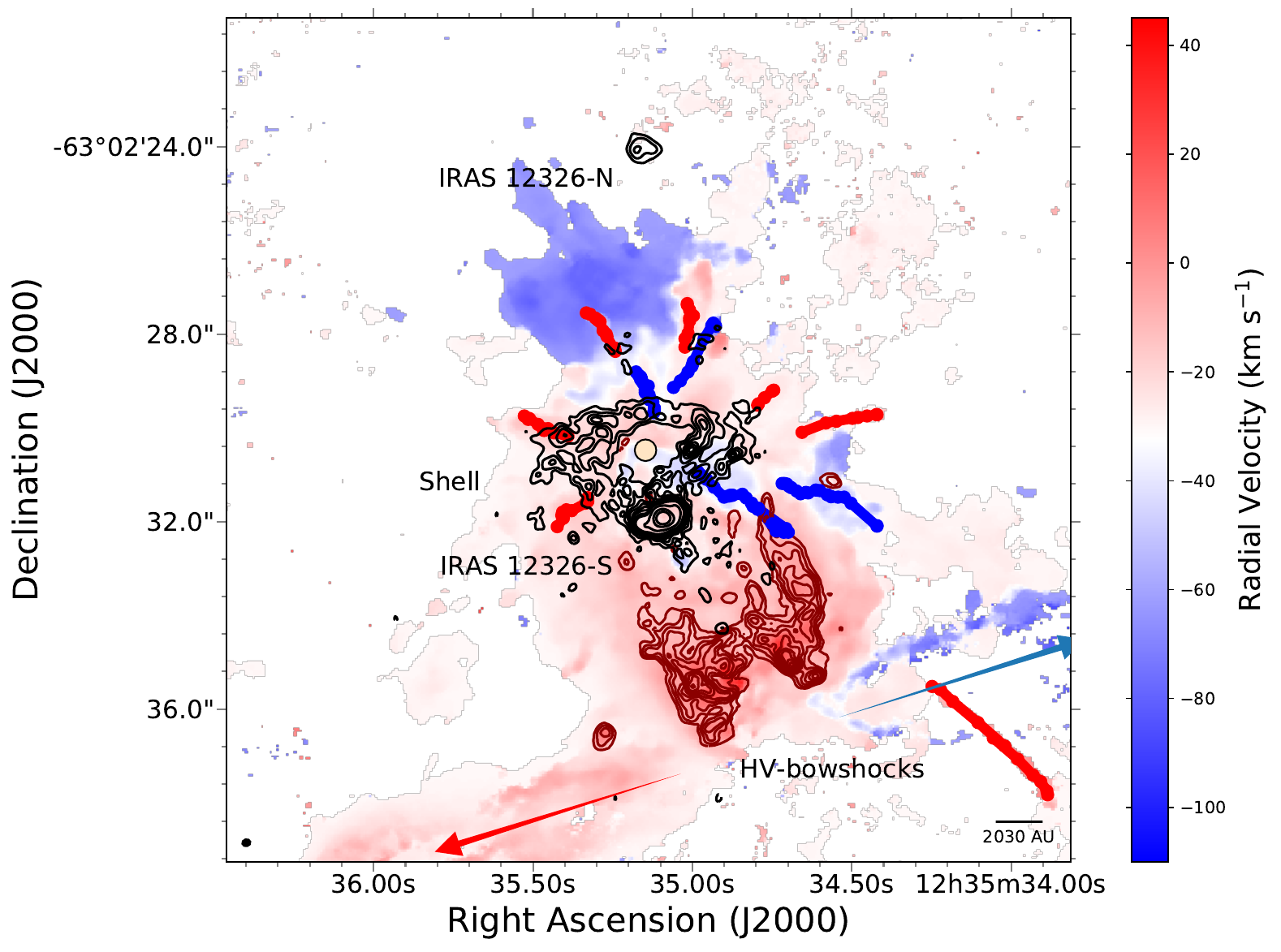}
\caption{ {\it Upper:} ALMA $^{12}$CO(2$-$1) moment zero map (grey scale) overlaid with the approaching (in blue colors) and receding (in red colors) explosive streamers 
discovered  in the IRAS12326$-$6245 molecular outflow, and the 1.3 mm continuum emission (black contours) arising also from this region. The IRAS12326$-$N source is 
evident in the north as a compact millimeter continuum source. The black contours are from $-$6, 6, 10, 15, 20, 25, 30, 40, 50, 100, 250,  and 
500 $\times$ 0.84 mJy beam$^{-1}$, the rms-{\it noise} level of the continuum image. Dark-red contours are tracing the redshifted extremely high velocity molecular 
gas ($+$70 to $+$2 km $^{-1}$) with a bow-shock morphology.  
The  contours range from 10\% to 100\% of the peak emission, in steps of 10\%.  The peak emission is 0.98 Jy beam$^{-1}$ km s$^{-1}$. In order to compute the 
moment zero map for the explosive outflow in IRAS12326$-$6245, we integrated the radial velocities from $-$120 to $+$80 km s$^{-1}$. On the other hand, we integrated for the
blueshifted streamers from $-$120 to $-$50 km s$^{-1}$, and $-$30 to $+$80 km s$^{-1}$ for the redshifted ones.  The systemic radial velocity is at $-$39.5 km s$^{-1}$ 
for the cloud in IRAS12326$-$6245 \citep{2021Duronea}. The bisque-color circle is tracing the position of the approximate center of the explosive outflow reported in this region.
  {\it Bottom:}  Same as above but with the $^{12}$CO(2$-$1) moment one map in color scales. The scale-bar on the right shows the radial velocities in km s$^{-1}$. 
  An electronic version of the CO(2$-$1) data cube is found here. }
\label{fig:Fig1}
\end{figure*}

\section{Observations} \label{sec:obs}

The ALMA observations were retrieved from \href{https://almascience.nrao.edu/aq/}{The Science Archive}. The observations were carried out during the Cycle 4 
science data program 2016.1.01347.S (PI: Silvia Leurini) in 2017 July 20 (with 41 antennas) and 23 (with 44 antennas). This science data program only included  
12 m diameter antennas with projected baselines ranging from 16.7 to 3700 m (12.8 to 2846 k$\lambda$).  The observations were pointed at the phase center located
in the sky position $\alpha_{J2000.0}$ = \dechms{12}{35}{35}{13}, and $\delta_{J2000.0}$ = \decdms{63}{02}{30}{8}.  IRAS 12326$-$6245 region was completely covered with a 
small mosaic of seven positions distributed in a Nyquist-sampled grid. The largest angular scale that can be recovered  with the present observations is 5.6$''$. 
The total integrated time on source was 66 min, distributed in the 7-pointing mosaic. 

The digital correlator was configured with eight spectral windows (SPW) centered at different frequencies in order to detected different molecular species at these millimeter wavebands. 
In this work, we concentrate on the SPW1 centered at 230.551 GHz, with a 244.141 kHz width, and divided into 1920 spectral channels, which resulted in a channel spacing 
of 317 m s$^{-1}$ across the SPW1. This spectral window was centered at this frequency to detect the $^{12}$CO(2$-$1) thermal line at a rest frequency of 230.5379700 GHz.  
We detect strong molecular CO line emission, see Figure~\ref{fig:Fig1}. We used the eight spectral windows to average line-free channels in order to obtain the continuum image.  
The continuum image is also presented in Figure~\ref{fig:Fig1}.

The weather conditions were very stable with an average precipitable water vapor of about 0.7 mm and an average system temperature of 100 K. 
 In order to reduce considerably the  atmospheric phase oscillation, during the observation  simultaneous observation of the 183 GHz water 
 line with water vapor radiometers was included.   Quasars J1107$-$4449, J1252$-$6737, and J1206$-$6138 were used to calibrate the bandpass, the atmosphere
  and the gain fluctuations, and the flux amplitude. Some of the quasar scans were repeated for different calibrations.
 
 To calibrate, image, and analyze the ALMA data we used the Common Astronomy Software Applications (CASA) package, Version 6.4.1.12.  
 Additionally, we also used some routines in Python to analyze the data \citep{astro2013}. The data was imaged using the routine \texttt{TCLEAN} with 
 the Robust parameter set equal to $+$0.5, a compromise between sensitivity and angular resolution.  
   
We obtained an image rms-noise for the continuum at 1.3 mm of 0.84 mJy beam$^{-1}$ at an angular resolution of  0.21$''$ $\times$ 0.18$''$ with 
a Position Angle (PA) of $-$68.2$^\circ$. We used a robust equals to 2.0 in \texttt{TCLEAN}.  This rms value is higher compared with the  ALMA theoretical rms-noise for this configuration, integration time, 
bandwidth, and frequency which is approximately 0.08 mJy beam$^{-1}$. A possibility to explain this, is the presence of a strong UCHII region localized 
within the field of view \citep{2000Henning} with an integrated flux of 1.8 Jy, which did not allow us to reach the theoretical noise level, see for an example \citet{2020Zapata}. 
Contrary to the continuum noises, we obtained a rms-noise for the line images of 2.1 mJy beam$^{-1}$ km s$^{-1}$  at an angular 
resolution of  0.18$''$ $\times$ 0.15$''$ with a PA of $-$63.0$^\circ$ (which translates into a brightness temperature of 1.277\,K). We used a robust equals to 0.5 in \texttt{TCLEAN}. 
We obtained similar values for the ALMA theoretical rms-noise for the spectral line images.

Using the continuum as a model, we undertook phase self-calibration, we then applied the acquired solutions to the spectral line channels. 
An improvement of a factor of about three in the rms-noise was obtained after applying self-calibration.

\begin{figure*}
\plotone{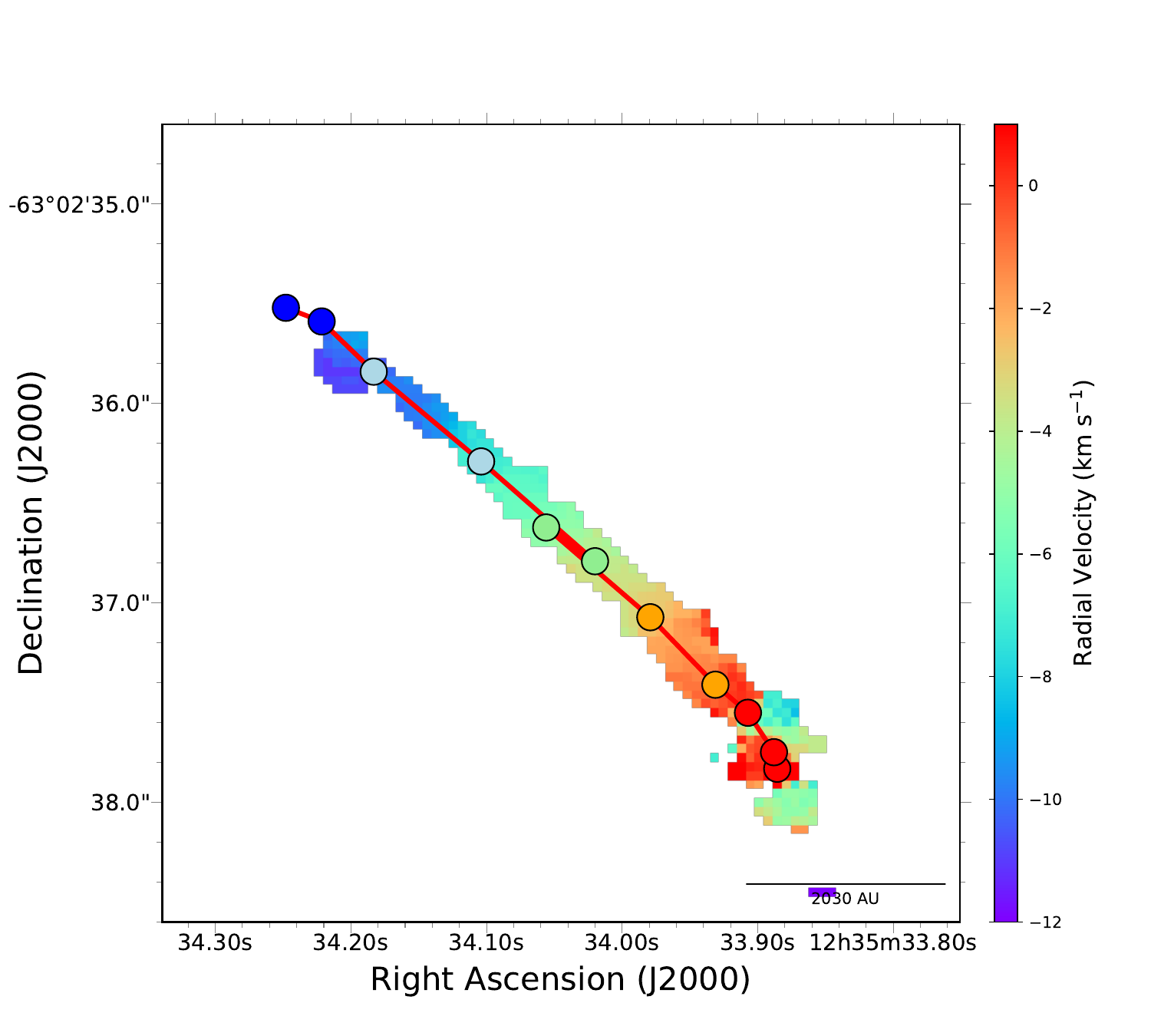}
\caption{ALMA moment one (color scale image) overlaid with the positions/radial velocities of the CO(2$-$1) condensations (color dots) from the most 
southern redshifted explosive molecular filament (see Figure~\ref{fig:Fig1}).  The filament and the condensations show a clear velocity gradient, with the most redshifted velocities
localized far from its origin, {\it i.e.} showing Hubble-Lemaitre like expansion motions. This velocity trend has been observed in other explosive molecular 
streamers, see for an example \citet{2017Zapata}. }
\label{fig:Fig2}
\end{figure*} 

\section{Results} \label{sec:results}

In Figure~\ref{fig:Fig1}, we show the results of the 1.3 mm continuum and $^{12}$CO(2$-$1) of the ALMA observations toward  the massive star forming regions IRAS12326$-$6245 and MC3-MIR2. 
In the continuum emission, we resolve the strong, spherical, and centrally peaked (sub)millimeter source detected with the Large APEX Bolometer Camera (LABOCA) on the APEX telescope \citep{2011dedes} 
towards IRAS12326$-$6245, and MC3-MIR2, into three compact sources that we name IRAS12326$-$N, the shell, and IRAS12326$-$S.   

Fitting a Gaussian to IRAS12326$-$N, we obtain an integrated flux of 138.1 $\pm$ 10 mJy, and a peak flux of 13.0 $\pm$ 0.9 mJy beam$^{-1}$. We also find that the 
deconvolved size results in a major axis of 685 $\pm$ 50 milli-arcsec, a minor axis of 547 $\pm$ 42 milli-arcsec, and a PA of 27$^\circ$ $\pm$ 14$^\circ$.  The centroid position in the sky for IRAS12326$-$N is     
$\alpha_{J2000.0}$ = \dechms{12}{35}{35}{155}, and $\delta_{J2000.0}$ = \decdms{63}{02}{24}{06}. The positional uncertainty for both the DEC and RA is about 0.02$''$.  
From Figure~\ref{fig:Fig1}, this source, in particular, seems to be double and possibly multiple. Future observations can resolve better its millimeter emission. This source is coincident with the
compact object G301.1364$-$00.2249 B, that based on its 4.8--8.6 GHz spectral index of +0.68, has been classified as a thermal radio jet candidate \citep{2012Guzman}, 
and the mid-infrarred source HLS2000 MIR 2 \citep{2000Henning} associated with a dusty object. We obtained a spectral index for IRAS12326$-$N of $+$1.9 from the SPWs 
centered at 216 and 230 GHz, likely associated with dust or optically thick emission from an HII region.   

The shell structure is first reported here (Figure~\ref{fig:Fig1}). This structure encircles the position in the sky  $\alpha_{J2000.0}$ = \dechms{12}{35}{35}{142}, 
and $\delta_{J2000.0}$ = \decdms{63}{02}{30}{64}, and has an average radius of 3$''$, though its shape is not exactly round or symmetric. Indeed it is more elongated in the 
East-West direction (4.5$''$) than in the North-South direction (about 2.7$''$). It consists of a series of $<0.5''$ (1000\,au) condensations, and some of its edges seem to comprise various 
threads or filament-like substructures (see for instance, two filaments within the western edge). The average width of the dusty shell is about 0.9$''$ (1800\,au).
The maximum peak emission from this structure is 30 mJy\,beam$^{-1}$, and a flux density of 1.3\,Jy, suggesting again that the shell is extended and faint. 
We estimated a spectral index for one condensation within the shell resulting in $+$3.1 from the SPWs mentioned before. This indicates that we are seeing optically thin dust emission. 
Thus a contribution of free-free emission should be minimal at these wavelengths.  We estimate the gas and dust mass in the shell by using its flux density and adopting the same assumptions 
as for the mass estimate of the dust shell in G5.89$-$0.39 \citep{2021Manuel}. That is, we assume optically thin emission, a grain opacity of 0.897 cm$^2$g$^{-1}$ (using the measured dust 
emissivity spectral index $\beta=1.1$) appropriate for dust with thin ice mantles at densities of $10^6$ cm$^{-3}$ \citep{1994Ossen}, a gas-to-dust ratio of 100, and a temperature ranging from 75--150\,K. 
With all of this, we obtain a shell mass ranging between 12.2 to 25.3\,M$_\odot$. Under these same assumptions and considering a 20\% contamination from free-free emission, we estimate 
a dust and gas mass for IRAS 12326$-$N between 1.0 and 2.1\,M$_\odot$.

Fitting again a Gaussian but now to IRAS12326$-$S, a strong and compact source just at the southern edge of the shell, see Figure~\ref{fig:Fig1},  
we obtain an integrated flux of 1.87 $\pm$ 0.02 Jy, and a peak flux of 1.00 $\pm$ 0.03 my beam$^{-1}$. We also find that the deconvolved size 
results in a major axis of 300 $\pm$ 10 milli-arcsec, a minor axis of 257 $\pm$ 5.0 milli-arcsec, and a PA of 101$^\circ$ $\pm$ 6$^\circ$.  
The position in the sky for IRAS12326$-$S is  $\alpha_{J2000.0}$ = \dechms{12}{35}{35}{091}, and $\delta_{J2000.0}$ = \decdms{63}{02}{31}{92}. 
The positional uncertainty for both the DEC and RA is about 0.003$''$ (this value is only statistical from the fitting). This source is coincident with the radio compact object G301.1364$-$00.2249 A, 
an optically thick UCHII with a spectral index of 2.0 \citep{2012Guzman,1998Walsh,2007Urquhart}. At millimeter wavelengths, we estimated a flat spectral index of $-$0.15 for this source,
which is consistent with an optically thin emission from the UCHII region. The rate of ionizing photons needed to maintain such a UCHII region is $\sim2.5\times10^{48}$ s$^{-1}$ \citep{1994kurtz}, 
which is compatible with a ZAMS star of spectral type O7.5-O8 \citep{1973nino,2005mar}. The position of this UCHII region is coincident with the dusty shell suggesting a probable relationship 
with the origin of the explosion as the sources Src I and BN in the Orion-KL region \citep{2011Zapata}. This may imply that the dusty shell, and the UCHII region were probably ejected with a similar acceleration. 
Some line analysis of the data covered by this ALMA project reveal an east-west velocity gradient ($\Delta$V=15 km s$^{-1}$) associated with the dusty shell, which we tentatively associated 
with an expansion motion. Similar expansion motions are observed in the dusty and ionized shells that surrounds the center of the explosive outflows, see Orion-KL, G5.89 or even DR21 
\citep{2009Zapata,2013Zapata,2020Zapata,2021Manuel}.  The dynamical age of the shell is estimated to be 700 yrs, in a very-well agreement with the estimated kinematic age of the explosive outflow.
 
In Figure~\ref{fig:Fig1}, we have also included the moment zero, and one maps of the $^{12}$CO(2$-$1) obtained with the present ALMA observations (grey and color scale images). 
In order to compute these maps, we integrated the radial velocities from $-$120 to $+$80 km s$^{-1}$.  We remark that with the present sensitive ALMA observations, we find very broad 
velocity gradient traced by the outflow located here ($\Delta$V $\sim$ 200 km s$^{-1}$). \citet{2000Henning} reported a velocity gradient for this same line of only $\Delta$V $\sim$80 km s$^{-1}$, 
more than a factor of two lower.  This can be explained by the great sensitivity of the ALMA observations that allows to detect fainter emission at higher velocities.   The CO emission is well resolved 
as one can see in Figure~\ref{fig:Fig1}.  The CO is tracing two main structures, one to the south, very elongated into an east-west orientation, likely associated with multiple classical bipolar outflows 
\citep[e.g.][]{2018Zapata,2015Zapata}, and the other one to the north with a morphology resembling an ellipsoidal shell, elongated along a direction with PA $\sim30\degr$, and centered 
on or near the position of IRAS~12326-S. The southern rim of this structure coincides with the northern ends of the redshifted bow-shocks shown in red contours in Figure \ref{fig:Fig1}. 
We do not find any evidence for an outflow energized by IRAS~12326-N. We also did not find any millimeter or even infrared (IRAC or MIPS in the Spitzer data archive) that could be related with 
bipolar east-west outflow.

\begin{figure*}
\plotone{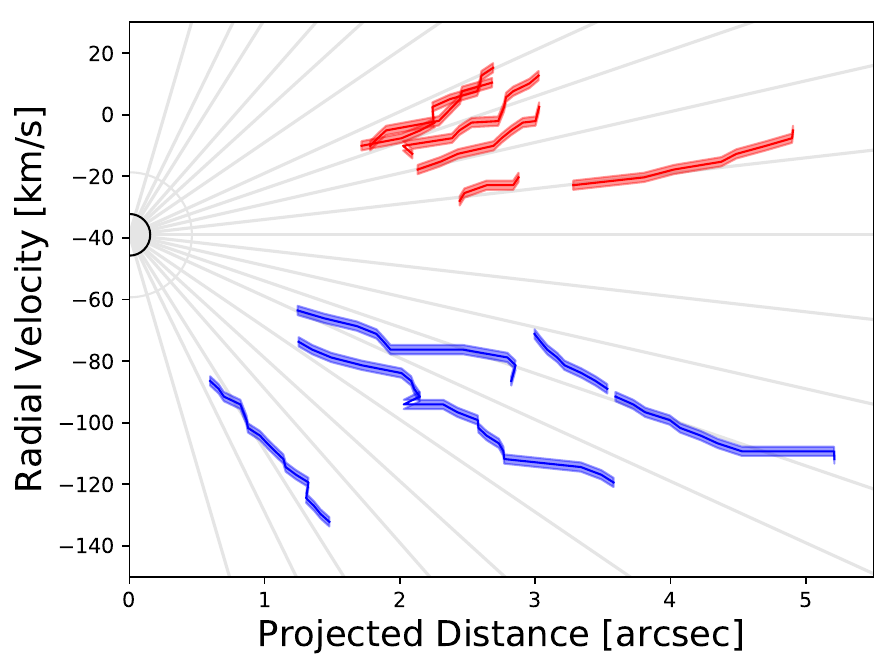}
\caption{Position$-$radial velocity digram of the explosive streamers detected in IRAS12326$-$6245 with our ALMA observations. 
The most southern and farther filament shown in Figure~\ref{fig:Fig1} is not plotted here for convenience.  We have plotted the blueshifted streamers in blue colors, 
while the redshifted ones in red. All the molecular streamers are mostly pointing to some place around a radial velocity to $-$39.5 km s$^{-1}$, the systemic cloud 
velocity of IRAS12326$-$6245.  In position, the center is located at $\alpha_{J2000.0}$ = \dechms{12}{35}{35}{147}, and $\delta_{J2000.0}$ = \decdms{63}{02}{30}{48}.  
In each filament the shaded or blurry blue and red colors represent the error area that is about 0.1$''$ for the on-the-sky distance.}
\label{fig:Fig3}
\end{figure*}

Figure~\ref{fig:Fig1} is centered intentionally in the shell-like outflow. In these images, we have additionally overlaid the 10 expansive streamers, and high velocity redshifted bow-shock 
structures emanating from a quasi-isotropic outflow reported here for the first time. Each of the streamers traces the position of one sequence of the CO condensations mapped at different radial 
velocities obtained from the velocity spectral cube, this procedure has been used already in many other studies, see for example \citet{2009Zapata,2013Zapata,2020Zapata,2022GuzmanCcolque}.  
In order to get a better view of this procedure, we have made additionally Figure ~\ref{fig:Fig2}. In this Figure, we show the moment one map of the CO(2$-$1) emission from the most southern redshifted 
expansive filament (see Figure~\ref{fig:Fig1}) overlaid with the positions of the molecular condensations obtained in the velocity spectral cube together with the radial velocities 
coded with a color-scale similar to that of the radial velocities in the moment one map. From this Figure~\ref{fig:Fig2}, it is clear to see the great correspondence between 
the positions of the condensations obtained in the spectral channel cube, and the moment one map. The velocity gradient in Figure 2 is very small ($\sim$10 km s$^{-1}$ over its 3$''$ length) 
compared to the full velocity range of the explosion.  Thus that given the large projected separation of this streamer from the suspected explosion center, it is likely that most of the motion is close to the plane of the sky.
The position on the sky were most of the expansive streamers seem to point is around the center of the dusty shell, at the position  $\alpha_{J2000.0}$ = \dechms{12}{35}{35}{147}, and $\delta_{J2000.0}$ = \decdms{63}{02}{30}{48} with a 
positional error of 2$''$. Unfortunately, we were not able to do a more complete statistical analysis to find the sky center of the explosion due to the low number of explosive streamers. 
In addition, south of IRAS12326-S we found no clear streamers, probably due to a more dense environment along this direction (which may hamper a clear kinematic and morphological 
footprint from the collimated streamers). Future sensitive ALMA observations will likely trace a more complete number of streamers, and the center might be refined. 
 
In Figure~\ref{fig:Fig3} is presented the radial-velocity vs. on-the-sky distance plot  of the 10 redshifted and blueshifted expanding streamers reported here. All the ten streamers follow more or less
straight lines and cluster to a range of radial velocities between $-$20 to $-$60 km s$^{-1}$, centered in the LSR systemic velocity of the cloud in IRAS12326$-$6245, which is $-$39.5 km s$^{-1}$ \citep{2021Duronea}. 
Additionally, the expanding streamers also follow Hubble-Lemaitre like expansion motions, that is, the radial velocities increase with the on-sky-position distances, 
noted already in many explosive outflows, e.g. \citet{2009Zapata,2013Zapata,2022GuzmanCcolque}. The physical deviations from straight lines are likely due to
 internal motions or sub structure of the expansive streamers.   There is not clear evidence of deceleration in the molecular streamers, likely because of its impulsive nature.  
 From this plot, we can estimate an approximate value for the kinematic time for the outflow, taking an average radial velocity of 40 km s$^{-1}$, and a distance of 3$''$ or $\sim$6000 au, 
 we obtain 700 yrs. This kinematic time for the outflow is consistent with the time (800 yr) obtained from the position most southern streamer (8$''$ away from the origin), 
 and the highest radial velocity observed in the streamers (100 km $^{-1}$).  In the appendix section, we have additionally included a PV diagram of the outflow caching some of the explosive 
 streamers with different orientations and radial velocities.       

Assuming that we are seeing optically thin CO(2$-$1) emission (more likely the emission is optically thick at these millimeter wavelengths, so this estimation should be considered a lower limit), 
and that we are in Local Thermodynamic Equilibrium (LTE), and following the equation 2 presented in \citet{2020Cortes}, we can estimate the mass for the explosive outflow located here.  
Taking a distance of 2.03 kpc, an excitation temperature of 30 K, a ratio between the molecular hydrogen, and the carbon monoxide of 10$^4$, 
and a range of velocities of 120 km s$^{-1}$,  we obtain a mass of 7 M$_\odot$, a momentum of 8$\times$10$^2$ M$_\odot$ km s$^{-1}$, and a kinetic energy of 10$^{48}$ erg.

\section{Discussion and Conclusions}

Our new ALMA observations lead us to propose that in IRAS12326$-$6245 there is an energetic explosive outflow associated with the shell-like outflow. 
All the morphological and kinematical features discussed in \citet{2017Zapata,2022GuzmanCcolque} are well distinguished for this outflow. 
The outflow consists of high-velocity straight narrow filament-like ejections with different orientations; the expansive streamers point back approximately to a common position; 
the radial velocities from the streamers follow a Hubble-Lemaitre like expansion; the outflow is embedded in a massive 5500\,M$_\odot$ gaseous condensation,
 and associated with a bright (10$^5$ L$_\odot$) infrared unresolved IRAS source. This outflow therefore adds one more case to the list of the explosive outflows 
 reported in Orion-KL, DR21, G5.89-0.39, IRAS 16076$-$5134, and S106. 

Taking these six explosive outflows reported in the literature (Orion-KL, DR21, G5.89-0.39, IRAS 16076$-$5134, IRAS 12326$-$6245, and S106), 
we can then estimate the rate of the explosive outflows in our Galaxy. Following \citet{2020Zapata,2022GuzmanCcolque}, we can update the rate of explosive outflows in the Galaxy. 
We adopt a time span of 15330 years between all 6 outflows (after taking into account their different distances to Earth), that all of them are immersed within a circle with a radius 2.8 kpc, 
and that the Galaxy is a thin disk with a radius of 15 kpc. With all of this, we estimate an updated rate of events of one every 90 years.   
This value is comparable to that reported by \citet{2020Zapata,2022GuzmanCcolque}. Our rough estimate will be refined once more explosive outflows will be detected.   
Moreover, the value reported here is well approximated to the rate of the Supernovae, 50 yrs \citep{2006Diehl}.    
Alternatively, we can compare this rate also to the massive star ($\geq$ 8 $M_\odot$)
formation rate, that can be estimated as follows. The global star formation rate in the Milky Way is $\sim$ 2 $M_\odot yr^{-1}$ \citep{Eli2022}. 
Since the average mass of stars in the initial mass function is $\sim$ 0.5 $M_\odot$ \citep{Pa2006}, we expect that 4 stars form per year.  Finally, since the fraction of massive stars is $\sim$0.5\% \citep{Pa2006} 
of the total number of stars, we get a massive star formation rate of 1 star every 50 yrs, that not surprisingly is very similar to the supernova formation rate.   

In conclusion, we report an explosive outflow localized in the high massive star forming region know as IRAS12326$-$6245.  We find over 10 molecular and collimated 
well-defined streamers, with Hubble-Lemaitre like expansion motions, and pointing right to the center of a dusty shell (reported for the first time here) localized in the northern 
part of the UCHII region known as MC3-MIR1 or UCHII G301.1A. The estimated kinematic age, mass, and energy for the explosion are $\sim$ 700 yrs, 7 M$_\odot$, and 10$^{48}$ erg, respectively.  
Therefore, IRAS12326$-$6245 adds a new case of explosive outflow as previously observed in Orion-KL, DR21, G5.89-0.39, IRAS 16076$-$5134, and S106.    

\begin{acknowledgments}
We would like to thank John Bally for his multiple helpful insights into this manuscript. 
L.A.Z. acknowledges financial support from CONACyT-280775 and UNAM-PAPIIT IN110618, and IN112323 grants, 
México. A.P. acknowledges financial support from the UNAM-PAPIIT IG100223 grant, and the CONAHCyT project 
number 86372 of the `Ciencia de Frontera 2019’ program, entitled `Citlalc\'oatl’, M\'exico.
This paper makes use of the following ALMA data: ADS/JAO.ALMA\#2016.1.01347.S ALMA is a partnership of ESO (representing its member states), 
NSF (USA) and NINS (Japan), together with NRC (Canada), MOST and ASIAA (Taiwan), and KASI (Republic of Korea), in cooperation with the Republic of Chile. 
The Joint ALMA Observatory is operated by ESO, AUI/NRAO and NAOJ.  The National Radio Astronomy Observatory is a facility of the National Science 
Foundation operated under cooperative agreement by Associated Universities, Inc.  This research has made use of the SIMBAD database, operated at CDS, Strasbourg, France.  
\end{acknowledgments}

\software{CASA (CASA Team et al. 2022) and astropy (The Astropy Collaboration 2013, 2018, 2022)}

\bibliography{ms}{}

\begin{thebibliography}{}
\expandafter\ifx\csname natexlab\endcsname\relax\def\natexlab#1{#1}\fi
\providecommand{\url}[1]{\href{#1}{#1}}
\providecommand{\dodoi}[1]{doi:~\href{http://doi.org/#1}{\nolinkurl{#1}}}
\providecommand{\doeprint}[1]{\href{http://ascl.net/#1}{\nolinkurl{http://ascl.net/#1}}}
\providecommand{\doarXiv}[1]{\href{https://arxiv.org/abs/#1}{\nolinkurl{https://arxiv.org/abs/#1}}}

\bibitem[{{Allen} \& {Burton}(1993)}]{1993Allen}
{Allen}, D.~A., \& {Burton}, M.~G. 1993, \nat, 363, 54,
  \dodoi{10.1038/363054a0}

\bibitem[{{Araya} {et~al.}(2005){Araya}, {Hofner}, {Kurtz}, {Bronfman}, \&
  {DeDeo}}]{2005Araya}
{Araya}, E., {Hofner}, P., {Kurtz}, S., {Bronfman}, L., \& {DeDeo}, S. 2005,
  \apjs, 157, 279, \dodoi{10.1086/427187}

\bibitem[{{Astropy Collaboration} {et~al.}(2013){Astropy Collaboration},
  {Robitaille}, {Tollerud}, {Greenfield}, {Droettboom}, {Bray}, {Aldcroft},
  {Davis}, {Ginsburg}, {Price-Whelan}, {Kerzendorf}, {Conley}, {Crighton},
  {Barbary}, {Muna}, {Ferguson}, {Grollier}, {Parikh}, {Nair}, {Unther},
  {Deil}, {Woillez}, {Conseil}, {Kramer}, {Turner}, {Singer}, {Fox}, {Weaver},
  {Zabalza}, {Edwards}, {Azalee Bostroem}, {Burke}, {Casey}, {Crawford},
  {Dencheva}, {Ely}, {Jenness}, {Labrie}, {Lim}, {Pierfederici}, {Pontzen},
  {Ptak}, {Refsdal}, {Servillat}, \& {Streicher}}]{astro2013}
{Astropy Collaboration}, {Robitaille}, T.~P., {Tollerud}, E.~J., {et~al.} 2013,
  \aap, 558, A33, \dodoi{10.1051/0004-6361/201322068}

\bibitem[{{Bally}(2016)}]{2016Bally}
{Bally}, J. 2016, \araa, 54, 491, \dodoi{10.1146/annurev-astro-081915-023341}

\bibitem[{{Bally} {et~al.}(2011){Bally}, {Cunningham}, {Moeckel}, {Burton},
  {Smith}, {Frank}, \& {Nordlund}}]{2011Bally}
{Bally}, J., {Cunningham}, N.~J., {Moeckel}, N., {et~al.} 2011, \apj, 727, 113,
  \dodoi{10.1088/0004-637X/727/2/113}

\bibitem[{{Bally} {et~al.}(2017){Bally}, {Ginsburg}, {Arce}, {Eisner},
  {Youngblood}, {Zapata}, \& {Zinnecker}}]{2017Bally}
{Bally}, J., {Ginsburg}, A., {Arce}, H., {et~al.} 2017, \apj, 837, 60,
  \dodoi{10.3847/1538-4357/aa5c8b}

\bibitem[{{Bally} \& {Zinnecker}(2005)}]{2005Bally}
{Bally}, J., \& {Zinnecker}, H. 2005, \aj, 129, 2281, \dodoi{10.1086/429098}

\bibitem[{{Bronfman} {et~al.}(1996){Bronfman}, {Nyman}, \&
  {May}}]{1996Bronfman}
{Bronfman}, L., {Nyman}, L.~A., \& {May}, J. 1996, \aaps, 115, 81

\bibitem[{{Cortes-Rangel} {et~al.}(2020){Cortes-Rangel}, {Zapata}, {Toal{\'a}},
  {Ho}, {Takahashi}, {Mesa-Delgado}, \& {Masqu{\'e}}}]{2020Cortes}
{Cortes-Rangel}, G., {Zapata}, L.~A., {Toal{\'a}}, J.~A., {et~al.} 2020, \aj,
  159, 62, \dodoi{10.3847/1538-3881/ab6295}

\bibitem[{{Dedes} {et~al.}(2011){Dedes}, {Leurini}, {Wyrowski}, {Schilke},
  {Menten}, {Thorwirth}, \& {Ott}}]{2011dedes}
{Dedes}, C., {Leurini}, S., {Wyrowski}, F., {et~al.} 2011, \aap, 526, A59,
  \dodoi{10.1051/0004-6361/200912874}

\bibitem[{{Diehl} {et~al.}(2006){Diehl}, {Halloin}, {Kretschmer}, {Lichti},
  {Sch{\"o}nfelder}, {Strong}, {von Kienlin}, {Wang}, {Jean}, {Kn{\"o}dlseder},
  {Roques}, {Weidenspointner}, {Schanne}, {Hartmann}, {Winkler}, \&
  {Wunderer}}]{2006Diehl}
{Diehl}, R., {Halloin}, H., {Kretschmer}, K., {et~al.} 2006, \nat, 439, 45,
  \dodoi{10.1038/nature04364}

\bibitem[{{Duronea} {et~al.}(2021){Duronea}, {Cichowolski}, {Bronfman},
  {Mendoza}, {Finger}, {Suad}, {Corti}, \& {Reynoso}}]{2021Duronea}
{Duronea}, N.~U., {Cichowolski}, S., {Bronfman}, L., {et~al.} 2021, \aap, 646,
  A103, \dodoi{10.1051/0004-6361/202039074}

\bibitem[{{Elia} {et~al.}(2022){Elia}, {Molinari}, {Schisano}, {Soler},
  {Merello}, {Russeil}, {Veneziani}, {Zavagno}, {Noriega-Crespo}, {Olmi},
  {Benedettini}, {Hennebelle}, {Klessen}, {Leurini}, {Paladini}, {Pezzuto},
  {Traficante}, {Eden}, {Martin}, {Sormani}, {Coletta}, {Colman}, {Plume},
  {Maruccia}, {Mininni}, \& {Liu}}]{Eli2022}
{Elia}, D., {Molinari}, S., {Schisano}, E., {et~al.} 2022, \apj, 941, 162,
  \dodoi{10.3847/1538-4357/aca27d}

\bibitem[{{Fa{\'u}ndez} {et~al.}(2004){Fa{\'u}ndez}, {Bronfman}, {Garay},
  {Chini}, {Nyman}, \& {May}}]{2004Faundez}
{Fa{\'u}ndez}, S., {Bronfman}, L., {Garay}, G., {et~al.} 2004, \aap, 426, 97,
  \dodoi{10.1051/0004-6361:20035755}

\bibitem[{{Fern{\'a}ndez-L{\'o}pez} {et~al.}(2021){Fern{\'a}ndez-L{\'o}pez},
  {Sanhueza}, {Zapata}, {Stephens}, {Hull}, {Zhang}, {Girart}, {Koch},
  {Cort{\'e}s}, {Silva}, {Tatematsu}, {Nakamura}, {Guzm{\'a}n}, {Nguyen Luong},
  {Guzm{\'a}n Ccolque}, {Tang}, \& {Chen}}]{2021Manuel}
{Fern{\'a}ndez-L{\'o}pez}, M., {Sanhueza}, P., {Zapata}, L.~A., {et~al.} 2021,
  \apj, 913, 29, \dodoi{10.3847/1538-4357/abf2b6}

\bibitem[{{G{\'o}mez} {et~al.}(2005){G{\'o}mez}, {Rodr{\'I}guez}, {Loinard},
  {Lizano}, {Poveda}, \& {Allen}}]{2005Gomez}
{G{\'o}mez}, L., {Rodr{\'I}guez}, L.~F., {Loinard}, L., {et~al.} 2005, \apj,
  635, 1166, \dodoi{10.1086/497958}

\bibitem[{{Guzm{\'a}n} {et~al.}(2012){Guzm{\'a}n}, {Garay}, {Brooks}, \&
  {Voronkov}}]{2012Guzman}
{Guzm{\'a}n}, A.~E., {Garay}, G., {Brooks}, K.~J., \& {Voronkov}, M.~A. 2012,
  \apj, 753, 51, \dodoi{10.1088/0004-637X/753/1/51}

\bibitem[{{Guzm{\'a}n Ccolque} {et~al.}(2022){Guzm{\'a}n Ccolque},
  {Fern{\'a}ndez-L{\'o}pez}, {Zapata}, \& {Baug}}]{2022GuzmanCcolque}
{Guzm{\'a}n Ccolque}, E., {Fern{\'a}ndez-L{\'o}pez}, M., {Zapata}, L.~A., \&
  {Baug}, T. 2022, \apj, 937, 51, \dodoi{10.3847/1538-4357/ac8c35}

\bibitem[{{Henning} {et~al.}(2000){Henning}, {Lapinov}, {Schreyer}, {Stecklum},
  \& {Zinchenko}}]{2000Henning}
{Henning}, T., {Lapinov}, A., {Schreyer}, K., {Stecklum}, B., \& {Zinchenko},
  I. 2000, \aap, 364, 613

\bibitem[{{Kurtz} {et~al.}(1994){Kurtz}, {Churchwell}, \& {Wood}}]{1994kurtz}
{Kurtz}, S., {Churchwell}, E., \& {Wood}, D.~O.~S. 1994, \apjs, 91, 659,
  \dodoi{10.1086/191952}

\bibitem[{{Loiseau} {et~al.}(2018){Loiseau}, {Pontalier}, {Milne}, {Goroshin},
  \& {Frost}}]{2018Loiseau}
{Loiseau}, J., {Pontalier}, Q., {Milne}, A.~M., {Goroshin}, S., \& {Frost},
  D.~L. 2018, Shock Waves, 28, 473, \dodoi{10.1007/s00193-018-0822-4}

\bibitem[{{MacLeod} {et~al.}(1998){MacLeod}, {Scalise}, {Saedt}, {Galt}, \&
  {Gaylard}}]{1998MacLeod}
{MacLeod}, G.~C., {Scalise}, Eugenio, J., {Saedt}, S., {Galt}, J.~A., \&
  {Gaylard}, M.~J. 1998, \aj, 116, 1897, \dodoi{10.1086/300538}

\bibitem[{{Martins} {et~al.}(2005){Martins}, {Schaerer}, \&
  {Hillier}}]{2005mar}
{Martins}, F., {Schaerer}, D., \& {Hillier}, D.~J. 2005, \aap, 436, 1049,
  \dodoi{10.1051/0004-6361:20042386}

\bibitem[{{Ossenkopf} \& {Henning}(1994)}]{1994Ossen}
{Ossenkopf}, V., \& {Henning}, T. 1994, \aap, 291, 943

\bibitem[{{Osterloh} {et~al.}(1997){Osterloh}, {Henning}, \&
  {Launhardt}}]{1997Osterloh}
{Osterloh}, M., {Henning}, T., \& {Launhardt}, R. 1997, \apjs, 110, 71,
  \dodoi{10.1086/312990}

\bibitem[{{Panagia}(1973)}]{1973nino}
{Panagia}, N. 1973, \aj, 78, 929, \dodoi{10.1086/111498}

\bibitem[{{Parravano} {et~al.}(2006){Parravano}, {McKee}, \&
  {Hollenbach}}]{Pa2006}
{Parravano}, A., {McKee}, C.~F., \& {Hollenbach}, D.~J. 2006, Revista Mexicana
  de Fisica Supplement, 52, 1

\bibitem[{{Raga} {et~al.}(2021){Raga}, {Rivera-Ortiz}, {Cant{\'o}},
  {Rodr{\'\i}guez-Gonz{\'a}lez}, \& {Castellanos-Ram{\'\i}rez}}]{2021Raga}
{Raga}, A.~C., {Rivera-Ortiz}, P.~R., {Cant{\'o}}, J.,
  {Rodr{\'\i}guez-Gonz{\'a}lez}, A., \& {Castellanos-Ram{\'\i}rez}, A. 2021,
  \mnras, 508, L74, \dodoi{10.1093/mnrasl/slab072}

\bibitem[{{Rodr{\'\i}guez} {et~al.}(2020){Rodr{\'\i}guez}, {Dzib}, {Zapata},
  {Lizano}, {Loinard}, {Menten}, \& {G{\'o}mez}}]{2020Rodriguez}
{Rodr{\'\i}guez}, L.~F., {Dzib}, S.~A., {Zapata}, L., {et~al.} 2020, \apj, 892,
  82, \dodoi{10.3847/1538-4357/ab7816}

\bibitem[{{Urquhart} {et~al.}(2007){Urquhart}, {Busfield}, {Hoare}, {Lumsden},
  {Clarke}, {Moore}, {Mottram}, \& {Oudmaijer}}]{2007Urquhart}
{Urquhart}, J.~S., {Busfield}, A.~L., {Hoare}, M.~G., {et~al.} 2007, \aap, 461,
  11, \dodoi{10.1051/0004-6361:20065837}

\bibitem[{{Walsh} {et~al.}(1998){Walsh}, {Burton}, {Hyland}, \&
  {Robinson}}]{1998Walsh}
{Walsh}, A.~J., {Burton}, M.~G., {Hyland}, A.~R., \& {Robinson}, G. 1998,
  \mnras, 301, 640, \dodoi{10.1046/j.1365-8711.1998.02014.x}

\bibitem[{{Zapata} {et~al.}(2018){Zapata}, {Fern{\'a}ndez-L{\'o}pez},
  {Rodr{\'\i}guez}, {Garay}, {Takahashi}, {Lee}, \&
  {Hern{\'a}ndez-G{\'o}mez}}]{2018Zapata}
{Zapata}, L.~A., {Fern{\'a}ndez-L{\'o}pez}, M., {Rodr{\'\i}guez}, L.~F.,
  {et~al.} 2018, \aj, 156, 239, \dodoi{10.3847/1538-3881/aae51e}

\bibitem[{{Zapata} {et~al.}(2015){Zapata}, {Lizano}, {Rodr{\'\i}guez}, {Ho},
  {Loinard}, {Fern{\'a}ndez-L{\'o}pez}, \& {Tafoya}}]{2015Zapata}
{Zapata}, L.~A., {Lizano}, S., {Rodr{\'\i}guez}, L.~F., {et~al.} 2015, \apj,
  798, 131, \dodoi{10.1088/0004-637X/798/2/131}

\bibitem[{{Zapata} {et~al.}(2011){Zapata}, {Loinard}, {Schmid-Burgk},
  {Rodr{\'\i}guez}, {Ho}, \& {Patel}}]{2011Zapata}
{Zapata}, L.~A., {Loinard}, L., {Schmid-Burgk}, J., {et~al.} 2011, \apjl, 726,
  L12, \dodoi{10.1088/2041-8205/726/1/L12}

\bibitem[{{Zapata} {et~al.}(2009){Zapata}, {Schmid-Burgk}, {Ho},
  {Rodr{\'\i}guez}, \& {Menten}}]{2009Zapata}
{Zapata}, L.~A., {Schmid-Burgk}, J., {Ho}, P. T.~P., {Rodr{\'\i}guez}, L.~F.,
  \& {Menten}, K.~M. 2009, \apjl, 704, L45, \dodoi{10.1088/0004-637X/704/1/L45}

\bibitem[{{Zapata} {et~al.}(2013){Zapata}, {Schmid-Burgk}, {P{\'e}rez-Goytia},
  {Ho}, {Rodr{\'\i}guez}, {Loinard}, \& {Cruz-Gonz{\'a}lez}}]{2013Zapata}
{Zapata}, L.~A., {Schmid-Burgk}, J., {P{\'e}rez-Goytia}, N., {et~al.} 2013,
  \apjl, 765, L29, \dodoi{10.1088/2041-8205/765/2/L29}

\bibitem[{{Zapata} {et~al.}(2017){Zapata}, {Schmid-Burgk}, {Rodr{\'\i}guez},
  {Palau}, \& {Loinard}}]{2017Zapata}
{Zapata}, L.~A., {Schmid-Burgk}, J., {Rodr{\'\i}guez}, L.~F., {Palau}, A., \&
  {Loinard}, L. 2017, \apj, 836, 133, \dodoi{10.3847/1538-4357/aa5b94}

\bibitem[{{Zapata} {et~al.}(2020){Zapata}, {Ho}, {Fern{\'a}ndez-L{\'o}pez},
  {Ccolque}, {Rodr{\'\i}guez}, {Reyes-Vald{\'e}s}, {Bally}, {Palau}, {Saito},
  {Sanhueza}, {Rivera-Ortiz}, \& {Rodr{\'\i}guez-Gonz{\'a}lez}}]{2020Zapata}
{Zapata}, L.~A., {Ho}, P. T.~P., {Fern{\'a}ndez-L{\'o}pez}, M., {et~al.} 2020,
  \apjl, 902, L47, \dodoi{10.3847/2041-8213/abbd3f}

\bibitem[{{Zinchenko} {et~al.}(2000){Zinchenko}, {Henkel}, \&
  {Mao}}]{2000Zinchenko}
{Zinchenko}, I., {Henkel}, C., \& {Mao}, R.~Q. 2000, \aap, 361, 1079,
  \dodoi{10.48550/arXiv.astro-ph/0007095}

\bibitem[{{Zinchenko} {et~al.}(1995){Zinchenko}, {Mattila}, \&
  {Toriseva}}]{1995Zinchenko}
{Zinchenko}, I., {Mattila}, K., \& {Toriseva}, M. 1995, \aaps, 111, 95

\end{thebibliography}
\bibliographystyle{aasjournal}

\begin{figure*}
\plotone{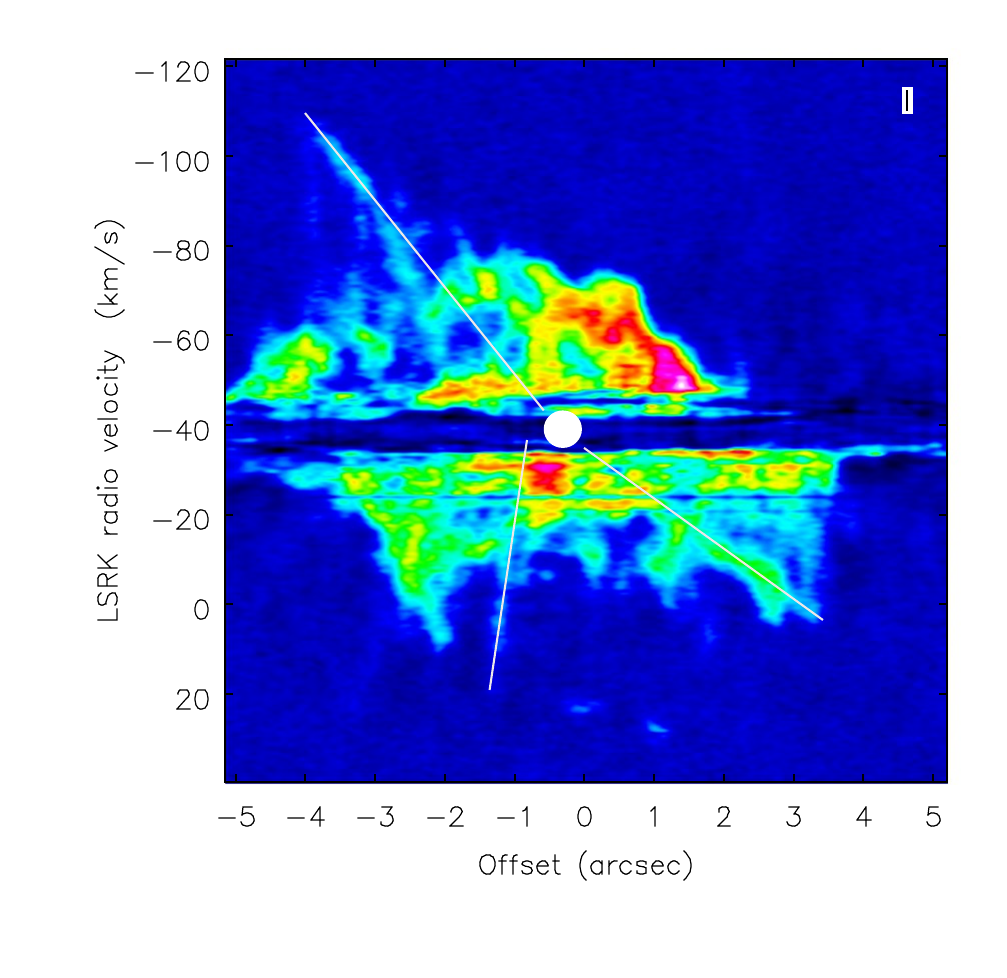}
\caption{Position$-$radial velocity digram of the explosive streamers detected in IRAS12326$-$6245 with our ALMA observations. The PV diagram is made at the position of the center of the explosion (see text), 
and a PA of 77.7$^\circ$. At this PA, we can elucidate three molecular streamers at high velocity. We have marked with white lines the streamers, and with a white circle the center of the explosion.    }
\label{fig:Fig4}
\end{figure*}

\appendix
In Figure~\ref{fig:Fig4}, we show a PV diagram of the ALMA CO(2$-$1) data. From this Figure, one can see some of the streamers having similar position angles directly from the ALMA data. 

\end{document}